\documentclass[fleqn,twoside]{article}

\usepackage[headings]{espcrc2}
\usepackage{graphics}
\newskip\humongous \humongous=0pt plus 1000pt minus 100pt
\def\caja{\mathsurround=0pt}
\def\eqalign#1{\,\vcenter{\openup1\jot \caja
       \ialign{\strut \hfil$\displaystyle{##}$&$
        \displaystyle{{}##}$\hfil\crcr#1\crcr}}\,}
\newif\ifdtup

\newcounter{eqnumber}[section]
\renewcommand{\theeqnumber}{\thesection.\arabic{eqnumber}}
\def\equn{\refstepcounter{eqnumber}
\eqno({\rm \theeqnumber})
}

\def\npb#1#2#3{{\rm Nucl. Phys. B}{\bf \ #1}, #3 (#2)}

\def\hepph#1{[hep-ph/#1]}

\newbox\charbox
\newbox\slabox
\def\s#1{{      % Feynman slash
        \setbox\charbox=\hbox{$#1$}
        \setbox\slabox=\hbox{$/$}
        \dimen\charbox=\ht\slabox
        \advance\dimen\charbox by -\dp\slabox
        \advance\dimen\charbox by -\ht\charbox
        \advance\dimen\charbox by \dp\charbox
        \divide\dimen\charbox by 2
        \raise-\dimen\charbox\hbox to \wd\charbox{\hss/\hss}
        \llap{$#1$}
}}

\def\spa#1.#2{\left\langle#1\,#2\right\rangle}
\def\spb#1.#2{\left[#1\,#2\right]}
\def\lor#1.#2{\left(#1\,#2\right)}

\catcode`@=11  % Make @ letter.

\def\Slash#1{\hskip 0.05 cm \slash\hskip -0.22 cm #1}

\def\eps{\epsilon}

\def\la{\langle}
\def\ra{\rangle}
\def\oneloop{{1 \mbox{-} \rm loop}}

\def\lsl{\not{\hbox{\kern-2.3pt $\ell$}}}
\def\ksl{\not{\hbox{\kern-2.3pt $k$}}}

\def\spa#1.#2{\left\langle#1\,#2\right\rangle}
\def\spb#1.#2{\left[#1\,#2\right]}
\def\lor#1.#2{\left(#1\,#2\right)}

\def\sand#1.#2.#3{%
  \left\langle\smash{#1}{\vphantom1}\right|{#2}%
  \left|\smash{#3}{\vphantom1}\right\rangle}
\def\sandp#1.#2.#3{%
  \left\langle\smash{#1}{\vphantom1}^{-}\right|{#2}%
  \left|\smash{#3}{\vphantom1}^{+}\right\rangle}
\def\sandpp#1.#2.#3{%
  \left\langle\smash{#1}{\vphantom1}^{+}\right|{#2}%
  \left|\smash{#3}{\vphantom1}^{+}\right\rangle}
\def\sandmm#1.#2.#3{%
  \left\langle\smash{#1}{\vphantom1}^{-}\right|{#2}%
  \left|\smash{#3}{\vphantom1}^{-}\right\rangle}
\def\sandpm#1.#2.#3{%
  \left\langle\smash{#1}{\vphantom1}^{+}\right|{#2}%
  \left|\smash{#3}{\vphantom1}^{-}\right\rangle}
\def\sandmp#1.#2.#3{%
  \left\langle\smash{#1}{\vphantom1}^{-}\right|{#2}%
  \left|\smash{#3}{\vphantom1}^{+}\right\rangle}

\def\Atree{A^{\rm tree}}

\def\Lz{\mathop{\hbox{\rm L}}\nolimits_0}
\def\Kz{\mathop{\hbox{\rm K}}\nolimits_0}

\def\BR#1#2{\la#1^+|\Slash{P}|#2^+\ra}

\def\tree{{\rm tree}}

\def\NeqFour{{\cal N} = 4}
\def\NeqOne{{\cal N} = 1}

\def\Fact{{\cal F}}

\title{Exploiting  Twistor Techniques for  One-loop QCD Amplitudes}

\author{
N.~E.~J.~Bjerrum-Bohr\address[UWS]{Department of Physics, 
University of Wales Swansea},
David~C.~Dunbar\addressmark[UWS]\address[comment]{Presented by David C. Dunbar at Loop and Legs 2006}
and
Harald~Ita\addressmark[UWS]  SWAT-06/465
}

\begin{document}

\maketitle

\section{Introduction}
Recently, a ``weak-weak'' duality between massless gauge theory and a
topological string theory propagating in twistor space has been
proposed in ref.~\cite{Witten:2003nn}. This duality implies surprising
structure within the $S$-matrix of gauge theories. In
supersymmetric theories this has been exploited to facilitate
considerable progress in computing scattering amplitudes. The
application of these ideas to QCD has taken longer, 
however recently progress has been made in developing techniques which 
can be applied to compute one-loop gluon scattering  
amplitudes~\cite{BDKrecursionA,LoopRecursionB,Forde:2005hh,Bern:2005hh}.
In this talk we discuss and review the work of ref.~\cite{Bern:2005hh}. 
We aim to establish recursion relations in the number of scattering gluons 
in an one loop amplitude.

%%%%%%%%%%%%%%%%%%%%%%%%%%%%%%%%%%%%%%%%%%%%%%%%%%%

\section{Twistor Inspired Techniques: Tree calculations} 
The link to twistor string theory is clearest if we express
amplitudes in terms of spinor variables by replacing the massless
momentum by $p_{a\dot a} = \lambda_a\bar\lambda_{\dot a}$ where
$p_{a\dot a}=(\sigma^\mu)_{a\dot a} p_\mu$ and use the spinor helicity
formalism~\cite{SpinorHelicity} for the polarisation vectors.
The BCFW on-shell recursion relations~\cite{Britto:2004ap} for tree
amplitudes are one of the remarkable formalisms which have arisen from the
duality. The recursion relations rely on the analytic structure of 
the amplitude after it has been continued to a function in 
the complex plane $A(z)$ by shifting the (spinorial) momentum 
of two reference legs,
$$
\lambda^1_a\longrightarrow \lambda^1_a+z\lambda^2_a \; , 
\;\; 
\bar\lambda^2_a\longrightarrow \bar\lambda^2_a-z\bar\lambda^1_a\,.
\equn$$
These shifts are equivalent to a shift in the momenta
\vskip -0.6 truecm 
$$
p^1_{a\dot a} \rightarrow 
p^1_{a\dot a} + z \lambda^1_a\bar\lambda^2_{\dot a} 
\,,\;  
p^2_{a\dot a} \rightarrow 
p^2_{a\dot a} - z \lambda^1_a\bar\lambda^2_{\dot a} \,.
\equn
$$
\vskip -0.1 truecm %By integrating $A(z)/z$ over a contour at infinity and 
%assuming $A(z) \rightarrow 0$ and relating this to the poles
%in the function $A(z)/z$, we can determine the unshifted amplitude
%$A(0)$.  
By integrating $A(z)/z$ over a contour at infinity and 
by assuming $A(z) \rightarrow 0$, the unshifted amplitude
$A(0)$ can be determined from the residues of the function 
$A(z)/z$.
The poles of this function are at $z=0$ and at $z_i$ given by
the factorisations of $A(z)$ when $P^2(z)=0$ for some
intermediate 
%momentum
propagator 
$i/P^2$, with the residue given by the product 
of the two tree amplitudes.  
A recursion relation is thus obtained which gives the 
$n$-point amplitude as a sum over %(shifted) 
lower point functions~\cite{Britto:2004ap}
$$
A(0)=  \sum _i \hat A_k(z_i)  \times { i \over P^2_i}  \times \hat A_{n-k+1}(z_i)\,.
\equn
$$ 
%In the above 
The summation only includes factorisations where the two
shifted legs $1$ and $2$ are 
on opposite sides of the pole.  
The tree amplitudes are evaluated at the value of $z$ such
that the shifted pole term vanishes, i.e. $P_i(z_i)^2=0$. 
%The analytic structure of the amplitude is the key ingredient in this process. 

The technique also extends %to many situations to generalised shifts
and in fact the correctness of the MHV formalism~\cite{CSW,CSWloop},
the other influential output from the twistor duality, can be derived
from this approach~\cite{Kasper,Bjerrum-Bohr:2005jr}.  The BCFW
recursion relations differ from the well established Berends-Giele
recursion relations~\cite{BerendsGiele} in that they are on-shell.

Although the duality relates string theory to $\NeqFour$
Super-Yang-Mills theory, the techniques inspired by the duality 
% seem to 
have much wider applicability. 
For gluonic tree amplitudes, this is
not so surprising since the tree amplitudes for gluonic scattering
coincide in QCD and $\NeqFour$ SYM. 
%However the techniques have been
%extended to a wide class of applications beyond those directly related
%to the duality. In particular, 
More surprisingly, the techniques may be applied to tree amplitudes with massive particles~\cite{MHVextensions} 
and to theories including gravity~\cite{BeBbDu,GravityBCF}.  
In retrospect, one may regard the duality as having been a tool which
has enabled the discovery of the techniques in field theory. 

In this talk we  %be intereste in 
discuss developing recursive techniques for one-loop
amplitudes. 

\section{One-Loop QCD Amplitudes}
A one-loop amplitude for massless particles can be 
expanded in the form
$$
A= \sum_i c_i I^i_4 
+\sum_i d_i I^i_3 
+\sum_i e_i I^i_2 
+R\,,
\equn
$$
where $I^i_n$ are scalar integral $n$-point integrals and 
$R$ denotes rational terms.
Loop amplitudes contain logarithmic (and dilogarithmic) terms which
would contain cuts in the complex plane when shifted.  Thus the entire
amplitude is not suitable for a recursion relation.
%However
%there are two ways an analytic recursion relation may be
%used:
However, recursion relations may be used on parts 
of the amplitude:

{\bf A)}  The rational terms 
%$R$ do not contain cuts so a recursion relation may be used for
$$
R \equiv   A- \Bigl( \sum_i c_i I^i_4 +\sum_i d_i I^i_3  +\sum_i e_i I^i_2  
\Bigr)\,.
\equn
$$

{\bf B)} The rational coefficients of the integral functions $c_i$, $d_i$ and
$e_i$.% are rational functions for which recursion might be possible.
 
The two approaches are complementary rather than competing.  
In both cases, to apply a recursion relation the key is an 
understanding of the singularity structure in the shifted coefficients %$A(z)$.
$R(z)$, $c_i(z)$, $d_i(z)$ or $e_i(z)$, which can be inferred from
the factorisation properties~\cite{BernChalmers} of the
full amplitude as $P^2 \rightarrow 0$,
{%\tiny
$$
%\hspace{-0.2cm}
\eqalign{
& A_{n}^{\oneloop}\
 \hskip -.12 cm
\ \ \ \longrightarrow^{\hspace{-0.7cm}P^2\rightarrow\,\, 0}\
\hskip .15 cm
\sum_{h=\pm}  \Biggl[
 A_{m+1}^{\oneloop} \,
            {i \over P^2 } \,
   A_{n-m+1}^{\tree} 
\cr
 % \hskip2.8cm \null  
 & + A_{m+1}^{\tree} \, {i\over P^2 } \,
   A_{n-m+1}^{\oneloop} 
%\label{LoopFact} 
%\cr
%& \hskip-1.5cm \null
 + A_{m+1}^{\tree} \, {i\over P^2} \,
   A_{n-m+1}^{\tree} \,
      \Fact_n \Biggr] \,.
\cr}\equn
\label{factorisation}
$$}
For the case of the coefficient $c_i(z)$
%From this our aim is to 
we obtain a recursion relation %for the coefficients
analogous to that for tree amplitudes,
$$
c_n(0) \; = \; \sum_{\alpha,h}  {A^h_{n-m_\alpha+1}(z_\alpha) \,
 {i\over P^2_{\alpha}}\, c^{-h}_{m_\alpha+1}(z_\alpha)} \,,
\equn\label{CoeffRecur}
$$
where $A^h_{n-m_\alpha +1}(z_\alpha)$ and $c^h_{n-m_\alpha+1}(z_\alpha)$ are
shifted tree amplitudes and coefficients evaluated at the residue
value $z_\alpha$\, and $h$ denotes the helicity of the intermediate state.

In order to have a valid bootstrap %we must have that 
the shifted coefficient has to vanishes as $|z|\rightarrow \infty$; 
otherwise there would be a dropped boundary term.  We can, however, 
impose criteria to prevent this from happening. Consider an integral 
and consider the unitarity cut which isolates the cluster on which 
the recursion will be performed, {\it i.e.} the one with the two 
shifted legs.

\begin{center}
\includegraphics{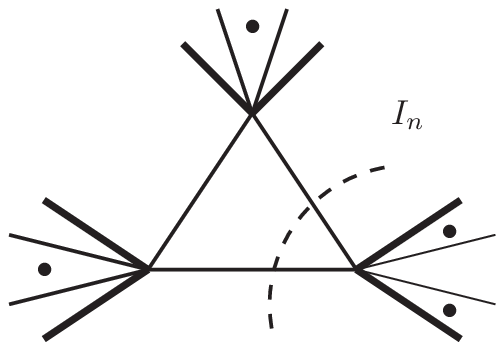}
\end{center}

\noindent
The dashed line in this figure indicates the cut. The recursion is to
be performed with the two shifted legs from the right-most
cluster. Then simple criteria for a valid recursion are:
\begin{enumerate}
\item The shifted tree amplitude on the side of the cluster
undergoing recursion vanishes as $|z| \rightarrow \infty$\,.
\item All loop momentum dependent kinematic poles
are unmodified by the shift. 
\end{enumerate}

Note that these are sufficient and not necessary conditions. 

\section{Complication: Spurious Singularities}
In addition to physical singularities, pieces of amplitudes also
contain {\it spurious singularities}. A spurious singularity is a
singularity that does not appear in the full amplitude but which
is present only in some parts of the amplitude. Typical examples 
are co-planar singularities such as
${ 1 \over \BR23  }$
which vanishes when $P=\alpha k_2 +\beta k_3$.  Such
singularities are common in the coefficients of integral functions.
These are not singularities of the full amplitude since, on the
singularity, the integral functions are not independent but combine to
cancel.  For example, for six-point kinematics, the product 
$\la 2 |\Slash{P}_{234} | 5 \ra$ 
vanishes when $t_{234}t_{612}-s_{34}s_{61}=0$.  At
this point the functions $\ln(s_{34}/t_{234})$ and
$\ln(s_{61}/t_{612})$ are no longer independent and the combination
$$
{ a_1 \over\la 2 | \Slash{P}_{234}  | 5 \ra } \ln(s_{34}/t_{234}) +
{a_2\over \la 2 | \Slash{P}_{234}  | 5 \ra }\ln(s_{61}/t_{612})\,,
\equn
$$ is non-singular provided that $a_1=a_2$ evaluated at the singularity.
Some spurious singularities can be controlled by the choice of basis
functions. For example expressions such as $\ln(r)/(1-r)^3$ will typically
appear in amplitudes where $r$ is the ratio of two momentum
invariants. These expressions have unphysical singularities 
at $r=1$ which cancel when combined with similar singularities 
in the rational terms. 
If we consider for a basis integral function the combination
$L_2(r)=( \ln(r)-(r-r^{-1})/2 )/(1-r)^3$ which
is finite as $r\longrightarrow 1$ then both the cut-constructible and
rational terms will be individually free of this spurious singularity.

\section{Supersymmetric Decomposition of QCD Amplitudes} 
In general we shall always examine color-decomposed amplitudes. 
Let $A_{n}^{[J]}$ denote the leading in color 
partial amplitude for gluon scattering due to an
(adjoint) particle of spin $J$\, in the loop. 
The three choices we are interested in are gluons ($J=1$), 
adjoint fermions ($J=1/2$) and adjoint scalars ($J=0$).  
It is considerably easier to calculate the contributions 
due to supersymmetric matter multiplets together with the 
complex scalar. The three types of supersymmetric multiplet are the
$\NeqFour$ multiplet and the $\NeqOne$ vector and matter
multiplets. We can obtain the amplitudes for QCD from the 
supersymmetric contributions via
$$
\eqalign{
A_{n}^{[1]} &\; = \; A_{n}^{\,\NeqFour}-4A_{n}^{\,\NeqOne\; {\rm chiral}}\;+\;A_{n}^{[0]}\,,
\cr
A_{n}^{[1/2]} &\; = \; A_{n}^{\,\NeqOne\; {\rm chiral}}\;-\;A_{n}^{[0]}\,.
\cr}
\equn
$$ 
The contribution from massless quark scattering can be obtained
from these trivially.  When we compute amplitudes in supersymmetric
theories we are calculating parts of the QCD amplitude - although the
process is incomplete unless we can obtain the non-supersymmetric
contribution $A_{n}^{[0]}$.

For $\NeqFour$ SYM, cancellations lead to considerable simplifications
in the loop momentum integrals. This is manifest in the ``string-based
approach'' of computing loop amplitudes~\cite{StringBased}.  As a
result, $\NeqFour$ one-loop amplitudes can be expressed simply as a
sum of scalar box-integral functions~\cite{BDDKa}. The
box-coefficients are ``cut-constructible''~\cite{BDDKa}. That is they
may be determined by an analysis of the cuts where the tree amplitudes
are the normal four dimensional one. This allows a variety of
techniques to be used in evaluating these.  Originally an analysis of
unitary cuts was used to determine the coefficients firstly for the
MHV case~\cite{BDDKa} and secondly for the remaining six-point
amplitudes~\cite{BDDKb}. Twistor inspired techniques, combined with
the application of cut-constructibility have been developed rapidly
over the past year~\cite{BeDeDiKo,Britto:2004nj,Bidder:2005in}.  For
theories with less supersymmetry the amplitudes are also
cut-constructible and, although more complicated, significant progress
has also been made for these
theories~\cite{Bidder:2004tx,BBDP,BrittoSQCD}. Consequently when we
wish to obtain QCD amplitudes, in many cases, the remaining component
is that for a scalar circulating in the loop.  In a numerical or
semi-numerical computation the scalar component is also the simplest
component to obtain~\cite{EllisEtAl}.  This piece is, in principle,
cut-constructible provided one performs cuts in exactly in the
dimensional regulating parameter~\cite{DimShift}; alternately one can
split into cut-constructible
pieces~\cite{BDDKb,Bern:2005hh,BrittoFengMastrolia} plus rational
terms and establish recursion relations for the
rational~\cite{BDKrecursionA}.

\section{Example: Split Helicity Amplitudes} 
As an example, let us consider the ``split helicity'' amplitude where
the negative helicity gluons in the colour-ordered amplitude are all
adjacent. 
$$
A_{n}^{\rm 1-loop}(1^-,2^-,\cdots, r^-, r+1^+,r+2^+,\cdots, n^+)\,.
$$ 
This helicity amplitude has several simplifying features and  
the tree amplitude in known from  the BCFW
techniques~\cite{SplitHelicity}. The $\NeqFour$ component of this
amplitude can most easily be obtained using unitarity so we
concentrate upon the other two components namely the $\NeqOne$ chiral
component and the scalar component.
These two amplitudes have several simplifications: firstly they
contain no box integral functions. The QCD amplitudes do contain such
functions but they are entirely determined by the $\NeqFour$
component. 

Consider a generic triangle or bubble integral function. Such a
function will contain at least one massive corner. The external legs
on this corner will be of split helicity.  (If the external legs on
this corner had the same helicity, then the internal legs would both
by necessity be of the opposite helicity. This tree amplitude vanishes
in $D=4$ for the scalar and fermionic states and does hence not 
contribute to the case we are considering.)  Let these legs be $a^-,\cdots
r^-,(r+1)^+,\cdots b^+$.  It can then be shown that the shift 
$$
\lambda_{r+1} \longrightarrow \lambda_{r+1}+z\lambda_{r}\,, \;\;\;
\bar\lambda_{r} \longrightarrow \bar\lambda_{r}-z\bar\lambda_{r+1}\,, 
\equn$$
satisfies the sufficiency conditions for a recursion relation.
Starting from the known five and six point functions we can then build
the result A) for the $\NeqOne$ chiral multiple and B) for the
cut-constructible part of the scalar contribution for a split helicity
partial amplitude. We present here the results for the case with
precisely three negative helicity gluons - the NMHV amplitudes. 

For the amplitudes with an $\NeqOne$ chiral multiplet running in the
loop the result is,
$$%\small
\hspace{-0.1cm}\eqalign{
& A_{n}^{\,\NeqOne\ {\rm chiral}}(1^-,2^-,3^-,4^+,5^+,\cdots, n^+)
 = 
\cr
&{\Atree \over 2} \Big[\! \Kz[ s_{n1} ]\! +\!\Kz[ s_{34} ]
\Big]
-{i \over 2}\Biggr[
\sum_{r=4}^{n-1} \,\hat d_{n,r}\,
{    \Lz [ t_{3,r} / t_{2,r} ] \over t_{2,r}  }
\cr
% &\hspace{-0.6cm}
&\!+\!
\sum_{r=4}^{n-2} \hat g_{n,r}
{    \Lz [ t_{2,r} / t_{2,r+1} ] \over t_{2,r+1}  }
\!+\!
\sum_{r=4}^{n-2} \hat h_{n,r}
{    \Lz [ t_{3,r} / t_{3,r+1} ] \over t_{3,r+1}  }
\Biggr]\cr}
\equn
$$
where, $\Lz(r)=\ln(r)/(1-r)$, $K_0[s]=-\ln(-s)+2+1/\eps$ and 
we use definitions $K_{a,b}\equiv k_a+k_{a+1}\cdots k_b$, $t_{a,b}=K_{a,b}^2$
and $Pr_{a,b}=\spa{a}.{a+1}\spa{a+1}.{a+2}\cdots \spa{b-1}.{b}$. 
The coefficients are given by 
$$%\small%
\hspace{-0.1cm}
\eqalign{
& \hat d_{n,r}\! =\!  
{  \la 3 | {K}_{3,r}{K}_{2,r}  | 1 \ra ^2
\la 3 | {K}_{3,r}\big[k_2,{K}_{2,r}\big]{K}_{2,r}  | 1 \ra
\over
 [ 2 | {K}_{2,r}  | r \ra
 [ 2 | {K}_{2,r}  | {r\!+\!1} \ra\, Pr_{3,r} % \spa3.4\ldots \spa{r-1} .{r}
Pr_{r+1,1} 
{t}_{2,r}\,t_{3,r}}\,,
\cr
&\hat g_{n,r}\! =\!  
\sum_{j=4}^{r}
{\la 3|K_{3,j} {K}_{2,j}|1\ra^2
\la 3|K_{3,j}{K}_{2,j}\!\big[k_{r\!+\!1},{K}_{2,r}\big]\!|1\ra
% \spa{j+3}.{j+4}
\over
[2|K_{2,j}|j\ra
[2|K_{2,j}|j\!+\!1\ra\,
Pr_{3,j}Pr_{j\!+\!1,1}   % {\spa{3}.{4}\spa4.5\ldots\spa{n}.1 
t_{3,j}t_{2,j}}\,, \cr
&\hat h_{n,r} = (-1)^n 
\hat g_{n,n-r+2}\bigl\vert_{(123..n)\to(321n..4)}\,.
}%\equn
$$
\noindent
The contribution for a scalar in the loop is~\cite{Bern:2005hh}
\def\Lzz{L_2}
{
$$
\eqalign{
&A_n^{[0]}(1^-,2^-,3^-,4^+,5^+,\cdots, n^+) = \frac{1}{3}\,A_{n}^{\,\NeqOne\, {\rm chiral}}
 \cr
%&\frac{1}{3}\,A_{n}^{\,\NeqOne\ {\rm chiral}}(1^-,2^-,3^-,4^+,5^+,\cdots, n^+) \cr
&%\hspace{2cm}
-\!{i \over 3}\Biggr[
\sum_{r=4}^{n-1} \hat d_{n,r}
{    \Lzz [ t_{3,r} / t_{2,r} ] \over t_{2,r}^3  }
\!+\!
\sum_{r=4}^{n-2} \hat g_{n,r}
{    \Lzz [ t_{2,r} / t_{2,r\!+\!1} ] \over t_{2,r\!+\!1}^3  }
\cr
&\hspace{2cm}\!+\!
\sum_{r=4}^{n-2} \hat h_{n,r}
{    \Lzz [ t_{3,r} / t_{3,r\!+\!1} ] \over t_{3,r\!+\!1}^3  }\Biggr]
%\cr &  \hskip 2 cm 
+ \hbox{rational}\,, \cr
}
\equn
$$}

\noindent 
where,
\def\Pr{Pr}
{$$%\small
\hspace{-0.1cm}
\eqalign{
&\hat d_{n,r}\! =\! 
{\!  \la 3 | {K}_{3,r}{k}_{2}  | 1 \ra 
  \la 3 | {k}_{2}{K}_{2,r}\!  | 1 \ra
\la 3 | {K}_{3,r}\!\big[k_2,\!{K}_{2,r}\!\big]\!{K}_{2,r}  | 1 \ra
\over
 [ 2 | {K}_{2,r}  | r \ra
 [ 2 | {K}_{2,r}  | {r\!+\!1} \ra\, 
\Pr_{3,5} %\spa3.4\ldots \spa{r-1}.{r}
\Pr_{r\!+\!1,1} % \spa{r\!+\!1}.{r\!+\!2}\ldots   
\spa{n}.1} ,
\cr
&\hat g_{n,r}\! =\!  \sum_{J=4}^{r} \biggl[ 
{
\la 3|K_{3,J} K_{2,J}P|1\ra
\la 3|K_{3,J} K_{2,J}\tilde P |1\ra
\over
[2|K_{2,J}|J\ra
[2|K_{2,J}|J\!+\!1\ra\,
}
\cr
& 
\hspace{2.7cm}\times {
\la 3|K_{3,J}K_{2,J}\big[k_{r+1},K_{2,r}\big]|1\ra
\over
\Pr_{3,J} \Pr_{J+1,1} %  \spa{3}.{4}\spa4.5\ldots\spa{n}.1 \,
t_{3,J}\,t_{2,J}  }  \biggr]\,,
\cr
&
\hat h_{n,r} =   
(-1)^n \hat  g_{n,n-r+2}\bigl\vert_{(123..n)\to(321n..4)}\,,
\cr}
\equn\label{ds0}
$$} 

\noindent
where $P=k_{r+1} K_{r+1,1}$ and $\tilde P=K_{r+1,1}k_{r+1}$. The
rational pieces of the scalar contribution remain to be calculated.

The formulas for the split helicity amplitudes with an arbitrary
number of negative helicity gluons are given
explicitly in ref.~\cite{Bern:2005hh}. 
%%%%%%%%%%%%%%%%%%%%%%%%%%%%%%%%%%%%%%%%%%%%%%%%%%
%

%%%%%%%%%%%%%%%%%%%%%%%%%%%%%%%%%%%%

\section{Conclusions}
The past two years have seen significant progress in the computation
of loop amplitudes in gauge theories. Although, these techniques have
arisen in the context of highly supersymmetric theories the process of
applying them to more general theories such as QCD is underway.

\end{document}